\documentstyle[12 pt]{article}
\begin{document}

\noindent \today,  SUBMITTED  TO JPHYSA.

\bigskip


\newcommand{\dsty}{ \displaystyle}

\newcommand{\ddt}{ \frac {\dsty d}{\dsty dt}}
\newcommand{\ba} {\begin{eqnarray}}
\newcommand{\eda} {\end{eqnarray}}
\newcommand{\beq} {\begin{equation}}
\newcommand{\edeq} {\end{equation}}
\def\rit{{\Bbb R}}
\def\cit{{\Bbb C}}
\def\nit{{\mathbb N}}
\def\zit{{\Bbb Z}}
\def\Ex#1{\langle#1\rangle}
\def\bra#1{\langle#1|}
\def\ket#1{|#1\rangle}

\newcommand\jump{\vspace{.5 cm}}
\newcommand\push{\setlength{\parindent}{2 cm} 
\setlength{\hangindent}{2 cm}}
\newcommand\petitpush{\setlength{\parindent}{1.1 cm} 
\setlength{\hangindent}{1.1 cm}}
\newcommand\ppush{\setlength{\parindent}{4 cm} 
\setlength{\hangindent}{4 cm}}
\newcommand\nopush{\setlength{\parindent}{0 cm} 
\setlength{\hangindent}{0.8 cm}}
\newcommand\refindent{\setlength{\parindent}{0.7 cm} 
\setlength{\hangindent}{3 cm}}
\newcommand\normal{\setlength{\parindent}{1.5 cm} 
\setlength{\hangindent}{0 cm}}

%
%
%
\def\footstrut{\baselineskip 12pt}
\hfuzz=5pt
\baselineskip 12pt plus 2pt minus 2pt
\vskip 24pt
\centerline{\bf FINITE DIMENSIONAL REPRESENTATIONS}
\vskip 12pt
\centerline{\bf
 OF THE QUADRATIC ALGEBRA:}
\vskip 12pt
\centerline{\bf  APPLICATIONS TO THE EXCLUSION PROCESS}
\vskip 36pt
\centerline{  K. Mallick$^{1,2}$  \ \ and \ \  S. Sandow$^{3}$}
\vskip 12pt
\centerline{ $^1$\it Service de Physique Th\'eorique, CE  Saclay}
\centerline{\it F--91191 Gif-sur-Yvette Cedex, France}
\vskip 12pt
\centerline{$^2$\it Laboratoire de Math\'ematiques UA 762 du
CNRS}
\centerline{\it DMI Ecole Normale Sup\'erieure}
\centerline{\it 45 rue d'Ulm, F--75231 Paris 05 Cedex, France}
\vskip 12pt
\centerline{$^3$  \it Department of Physics}
\centerline{\it  and Center for
                 Stochastic Processes in Science and Engineering}
\centerline{\it  Virginia Polytechnic Institute and State University}
\centerline{\it    Blacksburg, VA 24061-0435, USA  }
\vskip 36pt
 
 \noindent
 Short title :  QUADRATIC ALGEBRAS AND EXCLUSION MODEL.

\noindent
 PACS N$^o$:   02.10Sp 05.40  05.60

\centerline{\bf ABSTRACT}
We study the one dimensional
 partially asymmetric simple exclusion process
 (ASEP)  with open
boundaries, that describes a system of hard-core particles hopping 
stochastically on a chain coupled to reservoirs at both ends.
Derrida, Evans, Hakim and Pasquier 
[{\it J. Phys. A}  26, 1493 (1993)]  have shown that  
the stationary probability distribution of this
  model can be represented as a trace on 
a  quadratic algebra,
 closely related to
  the deformed oscillator-algebra. We construct
 {\it all finite dimensional} irreducible representations of this algebra.  
 This enables us to compute
 the stationary bulk density as well as all correlation
lengths for the ASEP  on a set of special curves of the phase diagram.


 \section{Introduction}
       The one dimensional  asymmetric simple exclusion
  process (ASEP)  has been extensively studied 
 by mathematicians \cite{Ligg}  and physicists
 (see \cite{DEPriv} and references therein) as one of the 
 simplest model of a system out of equilibrium.
 It is a diffusion model  with hard-core exclusion
    that can be used to describe  hopping  conductivity
 in superionic conductors, traffic flow,
 and interface growth; it can be mapped to  a lattice version
  of the Kardar-Parisi-Zhang equation \cite{KrugS}.

Subject to open boundaries, where particles are injected or extracted,
the ASEP displays a rather rich behaviour;
 it exhibits phase transitions in the
thermodynamic limit.  The exact steady state
for the fully asymmetric case 
 (particles jump only in one direction)
was computed in  \cite{ddm} for
special choices of the input parameters and more generally
in  \cite{DEHP} and \cite{sd}. 
These computations rely on  a recursion in the system size.
An elegant way to exploit the recursive property of the steady state is the matrix Ansatz
used in  \cite{DEHP}, where
   the  stationary probabilities are expressed as 
   matrix elements of products of  operators,
 which represent particles and holes.

  The same  technique was used
 in \cite{Sven} for the partially asymmetric
  exclusion process with open boundaries.
 The operators used here  generate a quadratic
 algebra   which is simply
 related to 
 the so-called deformed  oscillator-algebra.
 The phase diagram
 was derived in the thermodynamic limit, and the steady state current
   in each phase was computed. However,
   since suitable representations of the quadratic algebra were lacking
   general
 equal-time correlation functions
 in the stationary state could not be calculated.
 In a recent paper, Essler and Rittenberg \cite{Ritt}
    studied Fock representations  of the general quadratic algebra
  (that can generically be  mapped on the deformed oscillator algebra).
  They first found infinite dimensional matrices, and 
   then  gave conditions under which these infinite matrices have
 an invariant subspace
  of finite dimension. These constraints 
 were explicitely written in terms of the parameters of the
 partially ASEP
  in the case of a one dimensional 
 and a two dimensional invariant subspace.
  One and two dimensional  representations
   were constructed, 
   and  correlation functions  determined with the help
 of these representations.

      In this paper we use standard methods
     of linear algebra to classify and construct all irreducible
     finite dimensional representations of the deformed harmonic
     oscillator algebra, without having to start with infinite
      dimensional  matrices.  
We show that there is exactly one irreducible representation for any
  finite dimension $n$, 
 and  we  find   explicit
  constraints on the  parameters of the ASEP  that make this representation
  compatible with the boundary conditions (cf. Eq. (\ref{condition})).
These constraints define  the locus of points  in the phase diagram of the exclusion model that  
   are accessible by finite dimensional representations. In these regions
of the phase diagram  we
compute the bulk density in thermodynamic limit and
 all the correlation lengths. Our results prove some
  conjectures raised
  in \cite{Ritt}. Morever, 
to the best of our knowledge 
the finite dimensional representation presented here have not
 been reported in the 
literature.\footnote{ A reason for this fact may be that,
 as shown in \cite{DEHP},
 there are  no  irreducible
 representations of finite  dimension, strictly greater than 1,
 of  the algebra associated with the totally
 asymmetric exclusion process, 
  the  model   mostly studied. }
   The matrices  obtained  can also be useful in studying
 algebras associated with more general reaction-diffusion processes.

 This paper is organized as follows. In section 2, we recall
 the matrix  technique, relate it to the deformed harmonic oscillator
 algebra, and construct all
 irreducible finite dimensional representations of this algebra.
  Section 3 is devoted
  to the calculation of the bulk density and the correlation lengths in the thermodynamic limit
 in the regions of the phase space that are accessible by  finite
  dimensional representations. The last section concludes with some remarks on
possible generalizations.
Some general properties
 of the quadratic algebra which are shared by all
  representations are proved in  appendix A; these properties are used
 in appendix B to prove the non-triviality of the representations
 obtained in section 2.

\section{Finite dimensional representations of the quadratic algebra}

\subsection{Matrix Ansatz for the  ASEP with open boundaries}

   We consider the one dimensional partially asymmetric exclusion
 process with open boundary conditions. Each site  $i$ ($1\le i \le L)$
    of a one dimensional lattice of $L$ sites is either occupied
   by a particle $(\tau_i=1)$ or empty   $(\tau_i=0)$.
  The system evolves according to a stochastic dynamical rule:
   during each infinitesimal   time step $dt$, the  transitions
 allowed for the bond $(i,i+1)$ with $1\le i \le L-1$ are
 \begin{eqnarray}
   10  &\to& 01  \,\, \hbox { with rate }   1  \nonumber \\
   01  &\to& 10   \,\,  \hbox { with rate }   x  \, .
 \end{eqnarray}
  The parameter $x$ is positive and measures the strength of the driving field;
 one can assume with no restriction that $ x < 1$.
  The model studied here is not
  totally asymmetric, therefore $x > 0$.
 Particles
 are injected at sites 1 and $L$ with rates
  $(1-x)\alpha$ and $(1-x)\delta$ and extracted with rates  $(1-x)\gamma$
 and $(1-x)\beta $ respectively, where $\alpha,\beta,\gamma$ and $\delta$
  are strictly positive numbers.

 It was shown in \cite{DEHP} that the quantities
$f(\tau_1,\ldots,\tau_L)$ defined as
 \begin{eqnarray}
  f(\tau_1,\ldots,\tau_L) &=&  \langle W | \prod_{i=1}^{L}
  \left( \tau_i D  + ( 1 - \tau_i) E \right)   | V \rangle  \, 
  \label{matelem}
   \end{eqnarray}
   provide stationary solutions of the master equation of the ASEP if 
  the operators $D$, $E$ satisfy the algebra
\beq
 DE - x ED = (1-x)(D + E)  \, ,
\label{alg1}
 \edeq
 and the vectors $\langle W |$ and 
  $| V \rangle$ are such that
\ba
 (\beta D - \delta E) | V \rangle  = | V \rangle   \, , \nonumber \\
    \langle W | (\alpha E - \gamma D ) = \langle W | \, .
 \label{cbord}
 \eda
  The ASEP on a finite lattice has a unique stationary state
   (Perron-Frobenius theorem, see e.g. \cite{Gant}).
  Therefore,  all the quantities
  $f(\tau_1,\ldots,\tau_L)$  are proportional
 to the steady state probabilities of the system. If the matrix elements
  (\ref{matelem})  are not
 all equal to zero, then 
 the normalized stationary probabilities of
 the exclusion process are given by:
  \ba
P(\tau_1,\ldots,\tau_L) &=& \frac{1}{Z_L} f(\tau_1,\ldots,\tau_L)
  \, ,
\nonumber \\
  \hbox { with }  \,\,   Z_L  &=&  \langle W | ( D + E )^{L} | V \rangle
  \, .
   \end{eqnarray}
  To compute the stationary probability distribution
  of a system with $L$ sites, one needs to find representations
 of the algebra (\ref{alg1}) and of the boundary vectors 
 (\ref{cbord}). These representations must be such that the 
  matrix elements of length $L,$ i.e. of the type
 \beq
 \bra{W} D^{n_1}E^{m_1}  \ldots  D^{n_k}E^{m_k} \ket{V}  
  \hbox { with  } \,\, n_1+m_1+\ldots +n_k+m_k = L \, ,
\label{expres}
 \edeq
 do not identically vanish (i.e are non-trivial).
 In  appendix A, we find general properties,
  that a representation must possess  to ensure that matrix elements
 of the type (\ref{expres}) are non-trivial for a system of
 length $L.$

  In \cite{Ritt}, relations (\ref{alg1}) and  (\ref{cbord}) 
    are interpreted
  as a Fock representation of the quadratic algebra generated
 by the operators
 \beq
  A = \beta D - \delta E - 1   \,\, \hbox{ and } \,\,
   B = \alpha E - \gamma D - 1 \, .
\label{fock}
 \edeq
The matrices $A$ and $B$ act trivially on the boundary vectors:
 \beq
   A \ket{V}  = 0 \,\, \hbox{ and}  \,\, \bra{W}B = 0 \, .
  \label{vides}
  \edeq
 Thus $\ket{V}$ and $\bra{W}$ play the role
 of right and left vacuum for the operators $A$  and $B$.
  If $\alpha\beta \ne \gamma\delta$
   the operators $D$ and $E$ can be expressed as linear combinations
 of $A$ and $B$, which hence satisfy the quadratic algebra:
 \ba
 && (1-x)\alpha\gamma A^2 + (\alpha\beta - x \gamma\delta)AB
  + (\gamma\delta - x \alpha\beta)BA + (1-x)\delta\beta B^2
   =  \nonumber \\
&&(1-x)\left( (\alpha\beta -\gamma\delta)(\alpha+\gamma) -\alpha(\beta +\gamma)
  -\gamma (\alpha + \delta) \right) A 
  \nonumber \\
+ &&
 (1-x)\left( (\alpha\beta -\gamma\delta)(\beta +\delta) -\delta (\beta +\gamma)
  -\beta(\alpha +\delta) \right) B
\nonumber \\
 + &&
(1-x)\left( (\alpha\beta -\gamma\delta)(\alpha +\beta +\gamma +\delta)
  - (\alpha+\delta)(\beta+\gamma) \right) \, .
 \label{quadalg}
  \eda
    Conversely, a Fock representation of a general quadratic algebra
  can generically be transformed to a representation of an algebra
 of the type (\ref{alg1}) with boundary conditions
  like (\ref{cbord}).

\subsection{Classification of irreducible representations}

 We  show that the  $n$ dimensional irreducible representation of
  the algebra (\ref{alg1}) can be written  in a suitable basis as:
\begin{eqnarray}
\label{Drep}
    D & =& \left( \begin{array}{cccccc}
		  1+a &0&0&0&. &.\\
		  0&1+a x&0&0&& \\
		  0&0&1+a x^2&0&&\\
		   &&&.&&\\
		   &&&&.& \\
		   &&&&&1+a x^{n-1} 
		   \end{array}
		   \right) \\
  \hbox { and } \;\;\;
  E & = & \left( \begin{array}{cccccc}
		 1+\frac{1}{a }&0&0&0&. &.\\
		  1&1+\frac{1}{a x}&0&0&& \\
		  0&1& 1+\frac{1}{a x^2}&0&&\\
		   &&.&.&&\\
		   &&&.&.& \\
		   &&&&1&1+\frac{1}{ax^{n-1}}
		   \end{array}
                    \right) \,,
\label{Erep}
\end{eqnarray}
 $a$   being a non-zero  real parameter.

  In order to derive this result,
   we  start by defining  two operators $d$ and $e$ such that
 \beq
    D = 1 + d  \,\, \hbox{ and } \,\,  E = 1 + e \, .
 \edeq
    From  Eq. (\ref{alg1}),   we see that
  $d$ and $e$ satisfy the deformed harmonic oscillator
 algebra \cite{Bieder, Macfar}:
  \beq
    de  - x ed = 1 - x \, .
  \label{xdef}
  \edeq
  Using equation (\ref{xdef}), we note that 
  if $\lambda$ is an eigenvalue
 of the operator $de$, then
 ${\dsty \left( 1 + \frac{\lambda - 1}{x} \right)}$ is  an eigenvalue
  of the operator $ed$ (we recall that
 $x \ne 0$). In finite dimension the matrices
  $de$ and $ed$ have the same spectrum, which contains therefore
 the numbers:
 $$ \lambda, {\dsty 1 + \frac{\lambda - 1}{x}},
 {\dsty 1 + \frac{\lambda - 1}{x^2}},\ldots,
 {\dsty 1 + \frac{\lambda - 1}{x^k}}, \ldots k  \in {\bf N} \, . $$
 But there is only a finite number of distinct eigenvalues
 in a finite-dimensional representation: this implies that either
 $x$ is a root of unity or $\lambda=1$. Only the latter case is
 possible here: the  eigenvalues  of 
$de$ must all be equal to 1. The operator $de$ is invertible, and 
  so are  the matrices
 $d$ and $e$ (their determinant
 can not be zero). We rewrite now equation (\ref{xdef}) as:
 \beq
    d(e - d^{-1}) = x (e - d^{-1})d \, .
 \edeq
  Hence, our  problem is reduced to finding
  representations of the algebra 
  $$ {\cal D}{\cal E} = x {\cal E}{\cal D} \,\, \hbox{ with } 
 {\cal D} \hbox { invertible, } $$
 $$     \hbox{ where }  {\cal D} = d  \,\, \hbox{and } \,\, 
  {\cal E} = (e - d^{-1}) \, .$$
  If  $|a\rangle$ is an eigenvector of ${\cal D}$ 
 with eigenvalue $a$
 (which is different from 0 because  {\cal D} is invertible),
  then ${\cal E}|a\rangle$
 is an eigenvector 
 of ${\cal D}$ with eigenvalue $xa$ or is the null vector.
  Consequently, the space spanned by  the linearly independent vectors
 $\{ |a\rangle, {\cal E}|a\rangle,\ldots,  {\cal E}^k|a\rangle,\ldots \}$
 is stable under both ${\cal D}$ and ${\cal E}$. To obtain a  finite
  dimensional representation, there must be an integer $n$ such that
  ${\cal E}^n|a\rangle = 0.$ This integer $n$ is  also
  the dimension of the full representation space, since
  the representation  is irreducible.

  The proof 
  that the irreducible $n$ dimensional representation of the quadra-tic
 oscillator algebra is  given by (\ref{Drep}) and  (\ref{Erep}),
is completed by
  writing the matrices of 
  $D = 1+d = 1 +  {\cal D} $ and $E = 1+e = 1 + {\cal E} + {\cal D}^{-1}$
   in the basis
 $(|a\rangle, {\cal E}|a\rangle,\ldots,  {\cal E}^{n-1}|a\rangle) $.
 (Hereafter, this basis  will be  denoted  by
 $ (\ket{1},\ldots,\ket{n}).$)

\subsection{Boundary vectors and conditions on the rates }

      We construct   the boundary vectors $\ket{V}$ and $\bra{W}$
 associated with the irreducible representation
 of dimension $n$ found above. These boundary vectors are 
   right and left eigenvectors  of 
   the operators $A$ and $B$, defined by Eqs. (\ref{fock}), with 
   zero eigenvalue (see Eq. (\ref{vides})).
 In  the basis  $ (\ket{1},\ldots,\ket{n}) \,$ the matrices  $A$  and $B$ 
  are  bidiagonal, their eigenvalues are  readily obtained.
 One can check  that
   for $A$ and $B$ to have 0 as an eigenvalue,
 there must be two integers $k$ and $l$ between
 0 and $n-1$ such that:
 \ba
    \beta( 1 +  a x^k) - \delta ( 1 + \frac{1}{a x^k} ) - 1 &=& 0
   \nonumber  \\
   \hbox{and } \,\, 
  \alpha ( 1 + \frac{1}{a x^l} )  - \gamma ( 1 +  a x^l) - 1 &=& 0\, .
\label{syst1}
 \eda
 Necessarily $|l-k| = n-1$, otherwise a representation
  of dimension  less than $n$ would suffice.
  In the case $k=n-1$ and $l=0$, one has 
$$ | V \rangle  = | n\rangle  \,\,   \hbox{ and }  \,\,
   \langle W | = \langle  1 | \ . $$
 With such a choice, one has $\langle W| D^L |V \rangle  = 0$ for any $L.$
 This implies, using the definiteness property (proved in appendix A),
 that  all the quantities
 $f(\tau_1,\ldots,\tau_L)$ are equal to 0: this case has to be excluded.

 The only case that remains is
  $k=0$ and $l=n-1$;
    the  two  equations corresponding to (\ref{syst1}) are 
  \ba
  \beta a^2 + (\beta - \delta -1) a - \delta &=& 0 \nonumber \\
  \alpha \left( \frac{1}{ax^{n-1}} \right)^2 + (\alpha - \gamma - 1) 
\left(\frac{1}{ax^{n-1}}\right)  - \gamma &=& 0 \, ,
\label{syst2}
  \eda
  and they must have a common root $a$.
  Both equations have the same structure, their solutions can be written
  $ a = \kappa_{\pm}(\beta,\delta)$ and 
 $ \frac{1}{ax^{n-1}} = \kappa_{\pm}(\alpha,\gamma)$
 with
 \beq
\kappa_{\pm}(u,v) = \frac{ -u+v+1 \pm \sqrt{ (u-v-1)^2 + 4uv } }{2u} \,.
 \edeq
 This function already appeared in the study of
  the phase diagram  
of the partially asymmetric exclusion process  in \cite{Sven}.
 One can check that for $u$ and $v$ positive,
\ba
\kappa_{+}(u,v) > 0 \,\,  & \hbox{ and }  & -1 < \kappa_{-}(u,v) < 0 \, .
\label{ineqa} \\
 \hbox{ One also has } \,\,\,  \kappa_{+}(u,v)\kappa_{-}(u,v) & = &
  -\frac{v}{u} \, .
 \label{iden2}
\eda
  Therefore, the equations (\ref{syst2}) have a common root 
if and only if 
 \beq
 x^{1-n} =\kappa_{+}(\beta,\delta) \kappa_{+}(\alpha,\gamma)  \, .
\label{condition}
\edeq
Condition  (\ref{condition}) is an explicit constraint on the parameters
 of the ASEP. It
 defines the locus
 of points in the phase diagram accessible  to $n$-dimensional representations. 
 In  Fig. 1 are shown the 
  curves  (hyperbolae branches) in the
$\kappa_{+}(\alpha,\gamma)-\kappa_{+}(\beta,\delta)$ plane
where $n=1,..,8$ dimensional representations exist.
  It is worthwhile noticing that Eq. (\ref{condition}) 
proves a conjecture
made in  \cite{Ritt}, which states that finite dimensional
 representations exist only in the region  
$$\kappa_{+}(\beta,\delta) \kappa_{+}(\alpha,\gamma) > 1 \; .$$ 
If the condition (\ref{condition}) is fulfilled the common root
 of (\ref{syst2}) reads
\beq
\label{a}
a=\kappa_{+}(\beta,\delta) = \frac{1}{x^{n-1}\kappa_{+}(\alpha,\gamma)}
\;\;,\edeq
and there
 exist vectors
  $\bra{W}$  and $\ket{V}$ such that
 the  boundary conditions
(\ref{cbord}) are satisfied. These vectors can be computed as: 
\begin{eqnarray}
\label{vectors}
\langle W|&=&(w_1,w_2,...,w_n)\;\;\;,\;\;\;
|V>  = \left( \begin{array}{c}
		  v_1\\
		  v_2\\
		  .\\
		  .\\
		  .\\
		  v_n 
		   \end{array}
		   \right) \\
\label{vau}
\mbox {with }\;\;v_k&=&\prod_{m=1}^{k-1}\;
\frac{\delta  }
{\beta \;[1 - x^{-m}]\;
[a x^{m}-\kappa_-(\beta,\delta)]}\\
\label{we}
\mbox {and  }\;\;
w_k&=&\prod_{m=k-1}^{n-2}\;
\frac{- 1}
{[ 1  - x^{m-n+1} ]\;
[(a x^{m})^{-1}-\kappa_-(\alpha,\gamma)]}
\;\;\mbox {for}\;k=1,\ldots,n \, .\,\,\,\,\hskip 0.3cm
\;\; \end{eqnarray}
With this definition, one has $v_1=1$ and $w_n = 1 $.

  Hence, we have constructed an explicit 
 $n$-dimensional representation
(\ref{Drep}, \ref{Erep},  \ref{vectors}, \ref{vau} and \ref{we}) of the  
   quadratic algebra with the
 required boundary conditions, provided that the rates
   $\alpha,\beta,\gamma$ and $\delta$
  satisfy the constraint (\ref{condition}).
 As discussed earlier (section 2.1), 
we must verify
 that the matrix elements of length $L$ (\ref{expres}), computed
 with this representation, do not identically vanish.
 In appendix B, using general properties of the algebra 
derived in appendix A, we prove that these matrix elements  are non-trivial if 
the size $L$ of the system is larger than the dimension $n$ of the
 matrices. For $L<n$ there are certain choices of the parameters for which,
surprisingly enough,  all matrix elements of length
 $L$ vanish. We identify these choices of the parameters in appendix B.
In any case, our representation
 can be used to investigate thermodynamic behaviour.

\section{Bulk density and correlations in the thermodynamic limit}

In the thermodynamic limit the asymmetric exclusion process 
 exhibits three  different phases
in which the current
 and correlation functions  are given by different expressions.
 An exhaustive study has been carried out, mainly for the
  totally asymmetric model \cite{DEHP,sd,Derevans}. Much less is
 known about the partially asymmetric exclusion process: the phase
  diagram was obtained in \cite{Sven} from an exact calculation
  of the current in the thermodynamic limit. This phase diagram,
 shown in Fig.1, agrees with the mean-field prediction.
The bulk density has been computed in Ref. \cite{Ritt} in mean-field
approximation, and it was argued that the mean-field result is exact.
We recall here the description of the phase diagram together
 with some known results.

   The phases can be described as follows:
  \begin{itemize}
   \item Phase A (High Density): $\kappa_{+}(\beta,\delta) >
   \kappa_{+}(\alpha,\gamma)$  and  $\kappa_{+}(\beta,\delta) > 1$.

  In the thermodynamic limit, the current $J_A$ is
$$ J_A = (1-x) \frac {\kappa_{+}(\beta,\delta)}
 { \left[ 1 +  \kappa_{+}(\beta,\delta) \right]^2}\, , $$
 and the mean-field prediction for the density in the bulk is:
 $$ \rho_A^{MF} = 
\frac{\kappa_{+}(\beta,\delta)}{ 1 + \kappa_{+}(\beta,\delta)}
\, .$$

   \item Phase B (Low Density):
 $ \kappa_{+}(\alpha,\gamma) > \kappa_{+}(\beta,\delta)$ and 
  $ \kappa_{+}(\alpha,\gamma) > 1$.
  
    $$ J_B = (1-x) \frac { \kappa_{+}(\alpha,\gamma)}
  { \left[ 1 +  \kappa_{+}(\alpha,\gamma) \right]^2}\, , $$
    $$ \rho_B^{MF} = \frac{1}{1 +  \kappa_{+}(\alpha,\gamma)} \, . $$

     \item Phase C (Maximal Current): $\kappa_{+}(\beta,\delta) < 1 $
 and $\kappa_{+}(\alpha,\gamma) < 1 $.
$$ J_C = \frac{ 1 - x }{4}\, , $$  $$ \rho_C^{MF} = \frac{1}{2}  \, .$$
  \end{itemize}
 Phase A and B are separated by a coexistence line, defined by $\kappa_{+}(\beta,\delta)
 = \kappa_{+}(\alpha,\gamma) > 1$, through which the (mean-field)
 bulk  density is discontinuous. This phenomenon has been investigated
 in the totally asymmetric case, and reveals the presence of a shock
  between a region of low density and a region of high  
  density  \cite{DEHP,sd}.
  The same phenomenon happens in the partially asymmetric case, as
 will be confirmed by 
  our calculation of correlation functions.

     Little is known about general correlation functions
 of the partially asymmetric exclusion process. In \cite{Ritt},
 some correlation functions were computed with one and two dimensional
 representations, i.e. on lines 1 and 2 in Fig.1. 
   We shall now
identify
the generic structure of correlation functions for the case where $n$  
 dimensional representations exist. Such an exact result cannot
  be obtained from a mean-field analysis \cite{DEHP,Ritt}.
For example a two-point function is given by:
\begin{eqnarray}
\label{2-point}
<\tau_j \tau_k>&=& \frac {\langle W|C^{j-1}DC^{k-j-1}DC^{L-k}|V\rangle}
  {\langle W|C^L|V\rangle} \\
\mbox{with}\;\;C&=&D+E 
\;\;.\end{eqnarray}
The positions, on which a correlation function depends,
 enter above expression as an 
exponent of the matrix $C\:$. For this reason  we have to compute powers of 
the matrix $C$
which is given by:
\begin{eqnarray}
\label{C}
  C &= &D+E\; =\;\left( \begin{array}{cccccc}
		 \lambda_1&0&0&0&. &.\\
		  1&\lambda_2&0&0&& \\
		  0&1&\lambda_3&0&&\\
		   &&.&.&&\\
		   &&&.&.& \\
		   &&&&1&\lambda_n
		   \end{array}
                    \right) \\
\mbox{with}\;\;\lambda_k&=&2+a x^{k-1}+\frac{1}{a x^{k-1}}\nonumber\\
\label{lambda}
&=&2+x^{k-1}\;\kappa_+(\beta,\delta)\;+\;x^{n-k} \;\kappa_+(\alpha,\gamma)
\;\;.\end{eqnarray}
We have used Eqs. (\ref{condition}) and (\ref{a}) to derive the last line.

   For $L \gg 1$,  the dominating matrix elements of
 $C^L$ will be found in the invariant subspace of $C$ that corresponds
  to the largest eigenvalue. We must identify the highest eigenvalue,
 and examine whether $C$ can be diagonalized in the associated
  invariant space\footnote{We recall that the invariant space
 associated with the eigenvalue $\lambda$ is given by 
  $\bigcup_{i=0}^{\infty} \hbox{Ker} ( C - \lambda)^i \,.$}.
 We notice that all the eigenvalues of $C$ lie on the curve $ z \mapsto
 2 + z + 1/z \, ,$ with $z = a, ax,\ldots, a x^{n-1}\, .$ This implies that
  the largest eigenvalue of $C$ can only be  $\lambda_1$, or
 $\lambda_n$ or both of them if it happens that $\lambda_1 =\lambda_n\,.$
  We have:
 \ba
  \lambda_1 - \lambda_n &=& a + \frac{1}{a} - \left(
  a x^{n-1}+\frac{1}{a x^{n-1}} \right)
  =  ( 1 - x^{n-1}) \left( a - \frac{1}{a x^{n-1}} \right) \nonumber \\
  &=& ( 1 - x^{n-1}) (\kappa_+(\beta,\delta) - \kappa_+(\alpha,\gamma) ) 
 \eda
  There are three different cases to consider,
 that naturally correspond to different regions in the phase space:

\hfill\break
{\bf Bulk density and correlations in Phase A.} This phase is defined by 
$\kappa_{+}(\beta,\delta) >
   \kappa_{+}(\alpha,\gamma)$  and  $\kappa_{+}(\beta,\delta) > 1,$ 
 therefore $\lambda_1 > \lambda_n\,$. The leading eigenvalue of $C$
 is $\lambda_1$ and it is non degenerate. We write $\ket{\lambda_1}$
 (and $\bra{\lambda_1})$ for the corresponding right (and left) eigenvector.
 One should notice that $\bra{\lambda_1}$ is equal to $\bra{1}$.
 Hence we obtain that for large system sizes:
 \beq
  Z_L = \langle W|C^L|V\rangle \simeq \langle W|\lambda_1\rangle
  \Ex{\lambda_1 | V} \lambda_1^{L} \, .
 \label{zet_as}
  \edeq
 The bulk density is given by:
  \ba
  \rho_A = \Ex{\tau_i} &=& \frac {\langle W|C^{i-1}DC^{L-i}|V\rangle}
  {\langle W|C^L|V\rangle} \, \hbox{ with } \,  1 << i << L  \, ,
 \nonumber \\
    &\simeq& \frac{ \langle W|\lambda_1\rangle \lambda_1^{i-1}
   \Ex{\lambda_1 | D |\lambda_1}\lambda_1^{L-i}
   \Ex{\lambda_1 | V} } { \langle W|\lambda_1\rangle
  \Ex{\lambda_1 | V} \lambda_1^{L} } \nonumber \\
 &=& \frac{ 1 + a}{ \lambda_1}  \label{densiteA}\\
&=& \frac{\kappa_{+}(\beta,\delta)}{ 1 + \kappa_{+}(\beta,\delta)} \, .
\label{densiteA1} 
\eda
 Here we used that $\langle \lambda_1 | D = \langle \lambda_1 | ( 1 + a)$
 and $\Ex{\lambda_1 |\lambda_1 } = 1$ to derive the last two lines.
 This result agrees with the mean-field prediction, as it was conjectured in
 \cite{Ritt}.

  Next we compute $\langle W|C^{j-1}DC^{k-j-1}DC^{L-k}|V\rangle\;$.
  The result is 
a linear combination of powers of the eigenvalues. 
 Dividing it by the asymptotic
expression  (\ref{zet_as}) for $Z_L$ and using Eq. (\ref{2-point}) shows that
two-point function is a superposition of terms $(\lambda_i/\lambda_1)^m$, where 
$m$ is one of the distances $j,\;k-j,\;L-k\;$. Similarly, one can compute
higher order correlation functions. For a large system, i.e. for $L\gg n\;$,  
they decay exponentially with correlation lengths
\begin{eqnarray}
\xi_i&=&\left\{\;\ln \; \frac{\lambda_1}{\lambda_i} \;\right\}^{-1} \nonumber\\
&=&\left\{\; \ln\;\frac{2+x^{n-1}\;\kappa_+(\alpha,\gamma)\;+\;\kappa_+(\beta,\delta)}
{2+x^{i-1}\;\kappa_+(\beta,\delta)\;+\;x^{n-i} \;\kappa_+(\alpha,\gamma)
} \;\right\}^{-1}
\;\;i=1,...,n-1
\,.
\label{correlations A}
\end{eqnarray}
For $n=2$ the correlation length computed in \cite{Ritt} is recovered.
  One should noticed that the correlation lengths depend only
on the functions 
$\kappa_+(\beta,\delta)$ and $\kappa_+(\alpha,\gamma)$.

\vskip 1cm 
\hfill\break
{\bf Bulk density and correlations in Phase B.}
Phase  B, defined by $  \kappa_+(\alpha,\gamma) > 
 \kappa_+(\beta,\delta)\,$
and  $\kappa_+(\alpha,\gamma)>1\;$, is related to phase A
  by means of a symmetry: 
replacing $\tau_i$ by $1-\tau_{L+1-i}$
as well as  $\beta$ by $\alpha$, $\delta$ by $\gamma$ and
vice versa leaves the dynamics of the system invariant. We find that 
 the  bulk density  is:
 \begin{eqnarray}
 \rho_B &=& \frac{ 1 + ax^{n-1}} { \lambda_n} \label{densiteB} \\
&=& \frac{1}{1 +  \kappa_{+}(\alpha,\gamma)}  \,.
 \label{densiteB1}
  \end{eqnarray}
  And the correlation
lengths  are given by: 
 \begin{eqnarray}
\xi_i 
&=&\left\{\; \ln\;\frac{2+x^{n-1}\;\kappa_+(\beta,\delta)\;+\;\kappa_+(\alpha,\gamma)}
{2+x^{i-1}\;\kappa_+(\beta,\delta)\;+\;x^{n-i} \;\kappa_+(\alpha,\gamma)
} \;\right\}^{-1}
\;\;i=1,...,n-1
\,.
\label{correlations B}
\end{eqnarray}

\vskip 1cm 
\hfill\break
{\bf Correlation functions on the coexistence line.}
Phases A and B coexist
in a single system
on the line $\kappa_+(\beta,\delta) = \kappa_+(\alpha,\gamma)>1.$ 
 For the fully asymmetric process it was shown in 
\cite{ DEHP,sd}
that the density profile depends linearly on the position, which indicates  
that the border between the two phases can be anywhere with the same
probability. The same behaviour was observed in \cite{Ritt} for the
case where 2-dimensional representations exist, and  two-point correlations
 functions were  found to depend algebraically on the positions.

Let us discuss the case where $n$ dimensional representations exist. The eigenvalues
of $C$ are given by  Eq. (\ref{lambda}) which now reads 
\begin{eqnarray}
\label{lambda1}
\lambda_k=2\;+\;[\;x^{k-1}\;+\;x^{n-k}\;]\;\kappa_+(\beta,\delta)
\;\;\end{eqnarray}
because $\kappa_+(\alpha,\gamma) = \kappa_+(\beta,\delta)\;$.
  Obviously, these eigenvalues pairwise coincide: 
 $\lambda_k=\lambda_{n+1-k}$ for all $k\;$, and
 due to the line under the diagonal
 (see Eq. \ref{C})  $C$ is not 
diagonalizable. However, it can be transformed into a Jordan
 normal form with a two
dimensional Jordan block for each eigenvalue. This implies algebraic
    behaviour of the 
  correlation functions (see the discussion in Ref. \cite{hsp}).
  We denote by $(\ket{\lambda_1},\ket{\lambda_n})$ a basis of the
  2-dimensional invariant space associated with the highest eigenvalue
 $ \lambda = \lambda_1 = \lambda_n \, .$
 In this basis  the restriction of $C^L$ can be written as:
 \beq
     \lambda^{L-1} \left( \begin{array}{cc}
                    \lambda& 0 \\
                      L& \lambda    
                       \end{array}
                    \right)  \, . \\
 \edeq
 We choose $\ket{\lambda_n} = \ket{n}$
 and $\bra{\lambda_1} = \bra{1}$.
By decomposing on this normal basis, we find for large system sizes
\begin{eqnarray}
Z_L \;=\;\langle W|C^L|V\rangle \;\simeq
\;L \;\lambda^{L-1}\;\;,
\label{zet-as-coex}
\end{eqnarray}
because $w_n = v_1 = 1$.

  The  $m$-point correlation function can now be computed as 
\beq
\langle \tau_{x_1L} \tau_{x_2L}...\tau_{x_mL}\rangle
 = \frac {\langle W|C^{x_1L- x_0L -1}DC^{x_2L-x_1L-1}D\ldots
 DC^{L-x_mL}|V \rangle} {\langle W|C^L|V\rangle}
 \edeq
 $\hbox{with } 0=x_0 < x_1 < x_2 < \ldots < x_m < 1 \;.$
In the limit of
 large system size, we project on the $(\ket{\lambda_1},\ket{\lambda_n})$
 plane and keep only highest order terms:
 \ba 
&&\langle \tau_{x_1L} \tau_{x_2L}...\tau_{x_mL}\rangle 
  \nonumber  \\  
  &\simeq&
    \sum_{k=1}^m \frac{\langle V|C^{x_1L-1}D\ldots D |
\lambda_n \rangle\langle \lambda_n |C^{x_{k}L-x_{k-1}L - 1}| \lambda_1
   \rangle
    \langle \lambda_1|D \ldots DC^{L-x_mL}|V \rangle}
 {L \; \lambda^{L-1}} \nonumber \\
 &\simeq&  \sum_{k=1}^m  (x_{k}-x_{k-1}) \frac{ ( 1 + ax^{n-1})^{k -1}
 (1+a)^{m-k + 1}}
  {\lambda^m} \;.
\label{correlationC...}
\eda
To obtain the last line we used
 that
   $D |\lambda_n \rangle = (1 + ax^{n-1})\ket{\lambda_n}
  \,\, \hbox{ and }  \,\, \bra{\lambda_1} D = \bra{\lambda_1} ( 1 + a)\, .$
Using furthermore the expressions (\ref{densiteA}) and (\ref{densiteB}) 
for the bulk densities in phases A and B we find:
\ba
\langle \tau_{x_1L} \tau_{x_2L}...\tau_{x_mL}\rangle 
 \simeq &\sum_{k=1}^m  (x_{k}-x_{k-1}) \rho_B^{k-1} \rho_A^{m-k+1} \, .
 \label{correlationC}
 \eda
 The formula (\ref{correlationC}) clearly shows that on the coexistence
 line there is a shock between a low density phase $\rho_B$ and
 a high density phase $\rho_A$. This shock
 can be anywhere between $1$ and $L$  with the same probability.

\section{Conclusion }

   The finite dimensional Fock representations of the quadratic
 algebra, obtained here,  allow us to derive exact
 results for the partially asymmetric exclusion process
 on some special curves of the phase diagram. We believe that
 the formulae obtained  for the correlation functions
 are valid throughout phases A and B,
  but this is  a conjecture: in order to prove it, one would have to use
 infinite dimensional representations, that would lead to complicated
 calculations. Maybe a kind of `analytic continuation argument'
 could be used to obtain results valid on any curve of the form
$ \kappa_{+}(\beta,\delta) \kappa_{+}(\alpha,\gamma) = x^{1-n}$
 where $n$ is not necessarily an integer any more, but we don't
 know how to do that. An interesting, and related, question is
 to understand the intuitive physics behind condition (\ref{condition}):
  what makes a system, that can be analyzed with finite $n \times n$ matrices,
 different from the others?

    The matrices that we have constructed are also useful 
 in the representation theory of the algebras associated with
 more general reaction-diffusion processes. In \cite{Ritt}
 the following system of algebraic relations is derived
 for such algebras:
\ba
   \kappa_1 DE +    \kappa_2 ED  &=&  D + E  \, , \nonumber \\
    \kappa_3 D^2 &=& \kappa_4 DE + \kappa_5 ED   \, , \nonumber \\ 
     \kappa_6 E^2  &=& \kappa_7 DE  + \kappa_8 ED  \, ,
    \label{alggene}
\eda
 where the $\kappa_j$ can be computed from the rates of the
  reaction-diffusion process \cite{Ritt}.
           The representations of (\ref{alggene}) are a subset
 of those  of the deformed oscillator algebra. 
 An analysis, similar to the one given in section 2.3, would
  give conditions on the rates for a finite dimensional representation
 of (\ref{alggene}) to exist.
\vskip 1cm
{\bf \Large Acknowledgements}\\
It is a pleasure to thank C. Godr\`eche for many interesting discussions
  and comments. S. Sandow would like to acknowledge
the hospitality of the Service de Physique de l'\'Etat
 Condens\'e, CE  Saclay
as well as the  financial support
by the Deutsche Forschungsgemeinschaft.

\section{Appendix A }
\setcounter{equation}{0}
\def\theequation{A\arabic{equation}}

  In this appendix we show, without using any explicit representation,
   how to compute recursively  expressions of the type
 $$  \bra{W} D^{n_1}E^{m_1}  \ldots  D^{n_k}E^{m_k} \ket{V}  
  \hbox { with  } \,\, n_1+m_1+\ldots +n_k+m_k = L \, ,$$
 that we shall call
 {\it  matrix elements of length } $L$,
  and investigate conditions under which
 these quantities are different from zero.
 We need the following properties:

  \hfill\break  
 {\bf  Definiteness Property:} 
{\it  Matrix elements of length $L$ are either all strictly positive,
 or all strictly negative or all identically equal to zero.}

    Indeed, matrix elements of length $L$ are stationary
  solution of the master equation of the ASEP on a open chain 
  with $L$ sites  \cite{DEHP}. Therefore, according to the Perron-Frobenius
 theorem \cite{Gant}, they are unique 
  up to a multiplicative constant  and are all of the same sign.
  
\jump
\hfill\break
 {\bf  Reordering Property:} 
 {\it The expression
  $$  (D^{n_1}E^{m_1}  \ldots  D^{n_k}E^{m_k}) - 
 x^q E^{ {\sum_{i=1}^k m_i}}
  D^{ {\sum_{j=1}^k n_j}}  \, ,$$
 with    $L = n_1 + m_1 + \ldots +  n_k + m_k$ and 
 $ q = \sum_{j\le i \le k} n_j m_i $
 is equal to a linear combination
 of products of  $L-1$  operators $D$ or $E$ with strictly
 positive coefficients. }

  One begins with the formula
 (obtained by induction on $m_1$):
  $$ DE^{m_1} = x^{m_1}E^{m_1}D + (1-x)
  \left( \frac{1 - x^{m_1}}{1-x} E^{m_1}  + \sum_{k=0}^{m_1 -1}
  x^kE^kDE^{m_1 - 1 - k} \right) \, ,$$
  that allows to transfer a matrix $D$ from the left to the right
 of $E^{m_1}$,
 and proves the reordering property
 for $DE^{m_1} - x^{m_1}E^{m_1}D $.
 One concludes then by induction on $k$ and $n_k$.
 
\jump
\normal

The reordering property enables us to compute 
all matrix elements of length $L$ if 
expressions of  the type 
$\bra{W} E^{l} D^{L-l}  \ket{V}  \,\, \hbox{ with }\,\, 
   l=0,\ldots ,L\, $
and all matrix  elements of length $L-1$ are known.
Hence one can relate 
matrix elements 
of length $L$ to matrix elements of length $L-1$.
 This can be achieved starting with
  the following system which is a simple consequence of 
 (\ref{cbord}):
\ba
 \cases {  \alpha\bra{W} E^{l+1} D^{L-l-1}  \ket{V} 
  -\gamma\bra{W}  DE^{l} D^{L-l-1}  \ket{V}
  =  \bra{W} E^{l} D^{L-l-1}  \ket{V} \cr
  -\delta \bra{W}  E^{l} D^{L-l-1} E \ket{V}
 +\beta \bra{W} E^{l} D^{L-l}  \ket{V} 
 =  \bra{W} E^{l} D^{L-l-1}  \ket{V} \, \cr} \, .
 \eda
 This system  is  rewritten  according to the
 reordering property:
  \ba
  \cases{ \alpha\bra{W} E^{l+1} D^{L-l-1}  \ket{V} 
  -  x^{l}\gamma\bra{W} E^{l} D^{L-l}  \ket{V}
 =  {\cal W}_{L-1}  \cr 
  -x^{L-l-1}\delta \bra{W}  E^{l+1} D^{L-l-1} \ket{V}
 +\beta \bra{W} E^{l} D^{L-l}  \ket{V} 
 =  {\cal W'}_{L-1}  \cr}
\label{syst}
 \eda
 where $ {\cal W}_{L-1}$ and $ {\cal W'}_{L-1}$
 are positive linear combinations of 
  matrix element of length $L-1$.
 We must distinguish two cases:

  (i) if $\alpha\beta - x^{L-1}\gamma\delta \ne 0$, then  {\it  all
  expressions of length $L$ can be computed from
  expressions of length $L-1$.}
 Furthermore all matrix elements of length $L$ are equal to 0 if and only if
  all those of length $L-1$ are equal to zero.

  (ii) If $\alpha\beta - x^{L-1}\gamma\delta =0$, then
 {\it all matrix elements of length less or equal to $L-1$ 
 identically vanish}.
  Indeed, the l.h.s. of the equations in (\ref{syst})
 are proportional, therefore one must have:
  \ba
    0 =  \left | \begin{array}{cc}
                     \alpha &   {\cal W}_{L-1}  \\
           -x^{L-l-1}\delta &   {\cal W'}_{L-1} 
           \end{array}  \right| =  \alpha{\cal W'}_{L-1}
  + x^{L-l-1}\delta {\cal W}_{L-1}
  \eda
 Because of the definiteness property the positive linear combination
on the r.h.s.  can only be equal to 0, if
 all matrix elements of length $L-1$ are equal to 0.
 For any $n\le L-2$, one has 
$\alpha\beta - x^{n}\gamma\delta \ne 0$, because otherwise
 this would imply
$\alpha\beta=\gamma\delta = 0$ which is impossible since
 all the rates are strictly positive. Using 
  case (i), one  concludes that all 
 matrix elements of length less or equal
  to $L-1$ are identically 0.

\vskip 1cm

  In conclusion, we have proved the following properties of
 any representation
 of the algebra (\ref{alg1}) together with boundary conditions
 (\ref{cbord}): 

 \hfill\break
 {\bf  Property A: }  If there is one
 matrix element of length $L$ different from 0, then all
 matrix elements of length $\ge L$ are non-zero.

\hfill\break
  {\bf Property B:  }  Suppose that there is one
 matrix element of length $L$ different from 0,
 then for  expressions of length $< L$, one has to
 consider two cases:

   (i) If for all non negative integer $l$
  $\alpha\beta - x^{l}\gamma\delta \ne 0$, then all the
 expressions
$ \bra{W} D^{n_1}E^{m_1}  \ldots  D^{n_k}E^{m_k} \ket{V}$
   are non-zero for any system size.

    (ii)  If there exists one   $l$ 
  such that $\alpha\beta - x^{l}\gamma\delta = 0$, then all
 expressions of length $\le l$ identically vanish
 but  all   matrix elements of length $ > l$ are non-zero.

 \section{Appendix B}
\setcounter{equation}{0}
\def\theequation{B\arabic{equation}}

  We  first show, using Property A
 derived in Appendix A, that the representations found in
 sections 2.2 and 2.3 
  provide non-trivial weights  for systems of length
 larger than $n$ or equal to $n$. Indeed
 the matrix elements $\bra{W} D^l \ket{V} $ for $l = 0,\ldots, n-1$
  can not all vanish, otherwise one would have:
  \beq
   0 = \bra{W} D^l \ket{V} = \sum_{i=1}^{n} (\mu_i)^l w_iv_i \,\,
  \hbox{ for } \,\, l = 0,\ldots, n-1 \, ,
\label{Vander}
  \edeq
  where $\mu_i= 1 + ax^{i-1}$ is the $i$th eigenvalue of $D$.
 But the  relations (\ref{Vander}) can be interpreted
 as a system of $n$ equations with $n$  unknowns $(w_iv_i)$.
 This  system is a Van der Monde system, and as all the 
 eigenvalues $\mu_i$ are different from each other, the
 only solution is $(w_iv_i) = 0$
  for $i=1,\ldots, n$. But this is not the case:
 as one can see from the explicit formulae (\ref{vau} and \ref{we}),
  all  the components of $\ket{V}$ and $\bra{W}$  are different from 0.
  Consequently, the matrix elements  $\bra{W} D^l \ket{V} $
 cannot vanish for all $l$, i.e. there is an  $l' < n$
 such that   $\bra{W} D^{l'} \ket{V} \ne 0$. 
 According to Property A, of appendix A, we conclude that
 the representation (\ref{Drep}) and (\ref{Erep}), together with
 the boundary vectors  given by (\ref{vectors}) and
 (\ref{vau})  always provides the stationary probabilities
 of the partially asymmetric exclusion process for systems
 of size bigger than $n$. 

 \vskip 0.5 cm

 Secondly, for systems of  size  strictly less than $n$, one has to check,
 according to Property B of  appendix A, whether
  $\alpha\beta -  x^{k} \gamma\delta$  can be 
 equal to 0 
 for some non-negative integer $k$. Using (\ref{iden2}) and (\ref{condition}),
this condition is equivalent to:
 \beq
    x^{n-1-k} =\kappa_{-}(\beta,\delta) \kappa_{-}(\alpha,\gamma) \, .
\label{cond2}
\edeq
 So, if (\ref{cond2}) is never satisfied, the $n$-dimensional
 representation found above allows to compute the  stationary weights
 for any system size.  If there is a 
 $k_0$, such that (\ref{cond2}) is true, then $k_0$ has to be less
 than $n-2$ (because of \ref{ineqa}),  and the $n$-dimensional
 representation provides non-trivial weights only for systems of size
 bigger or equal to $k_0 + 1 $.

\newpage

{\bf \Large Caption to the figure}
\vskip 12pt
\noindent
Fig.1: Phase diagram for the ASEP with $0<x<1$ in terms of 
$\kappa_{+}(\alpha,\gamma)$ and 
$\kappa_{+}(\beta,\delta)$ where
$\kappa_{+}(u,v) = \frac{1}{2u}[-u+v+1 \pm \sqrt{ (u-v-1)^2 + 4uv}]$.
The phases are separated by the solid lines. The dashed lines are the lines
where finite dimensional representations exist
 for $x = 3/4,$ and  the numbers attached to these
 lines are the dimensions of the corresponding representations.  

\vskip 2cm





\begin{thebibliography}{20}


 \bibitem{Ligg}  Liggett TM: {\it  Interacting Particle Systems}
NY: Springer Verlag (1985)
 
\bibitem{DEPriv} Derrida B  and Evans MR: The asymmetric
  exclusion model: exact results through a matrix approach
  {\it in Nonequilibrium Statistical Mechanics in One Dimension
 Ed. Privman V.}  Cambridge UK: C.U.P (1997)

  \bibitem{KrugS} Krug J and Spohn H: Kinetic Roughening of Growing
Surfaces { \it in Solids far from Equilibrium;
Ed  Godr\`{e}che C, }  Cambridge UK: C.U.P (1991)

\bibitem{ddm} Derrida D, Domany E and Mukamel D: An Exact Solution of a
One-Dimensional Asymmetric Exclusion Model with Open Boundaries
{\it J. Stat. Phys.} 69, 667 (1992)

\bibitem{DEHP} Derrida B, Evans MR, Hakim V, Pasquier V:
Exact Solution of a 1d  Asymmetric Exclusion Model using
a Matrix Formulation {\it J. Phys. A }  26, 1493 (1993)

 \bibitem{sd} 
Sch\"utz G and Domany E:  Phase transitions in an exactly
  soluble one-dimensional exclusion process
{\it J. Stat. Phys.} 72, 277 (1993)


\bibitem{Sven} Sandow S: Partially asymmetric exclusion process
with open boundaries {\it Phys. Rev.} E50, 2660 (1994)

 \bibitem{Ritt} Essler FHL and Rittenberg V: Representations of 
the quadratic algebra and partially
asymmetric diffusion with open boundaries {\it  J. Phys. A }
  29, 3375 (1996)

  \bibitem{Gant}  Gantmacher FR: {\it Application of the Theory
  of Matrices,} Interscience (1959)

\bibitem{Bieder} Biedenharn LC: The Quantum Group $SU(2)_q$
  and a $q$-analogue of the Boson Operators {\it J. Phys. A } 22, L873 (1989)

\bibitem{Macfar}  Macfarlane AJ: On $q$-analogues of the Quantum Harmonic
  Oscillator and the Quantum Group $SU(2)_q$ {\it J. Phys. A} 
 4581 (1989)
 
 \bibitem{Derevans} Derrida B and Evans MR: Exact correlation functions
 in an asymmetric exclusion model with open boundaries
 {\it J. Physique I (France)} vol.3, 311 (1993)
 

 \bibitem{hsp}
	 Hinrichsen H,  Sandow S and Peschel I:
  On matrix product ground states for reaction-diffusion models
		{\it J. Phys. A: Math. Gen.} 29  2643  (1996)


\end{thebibliography}
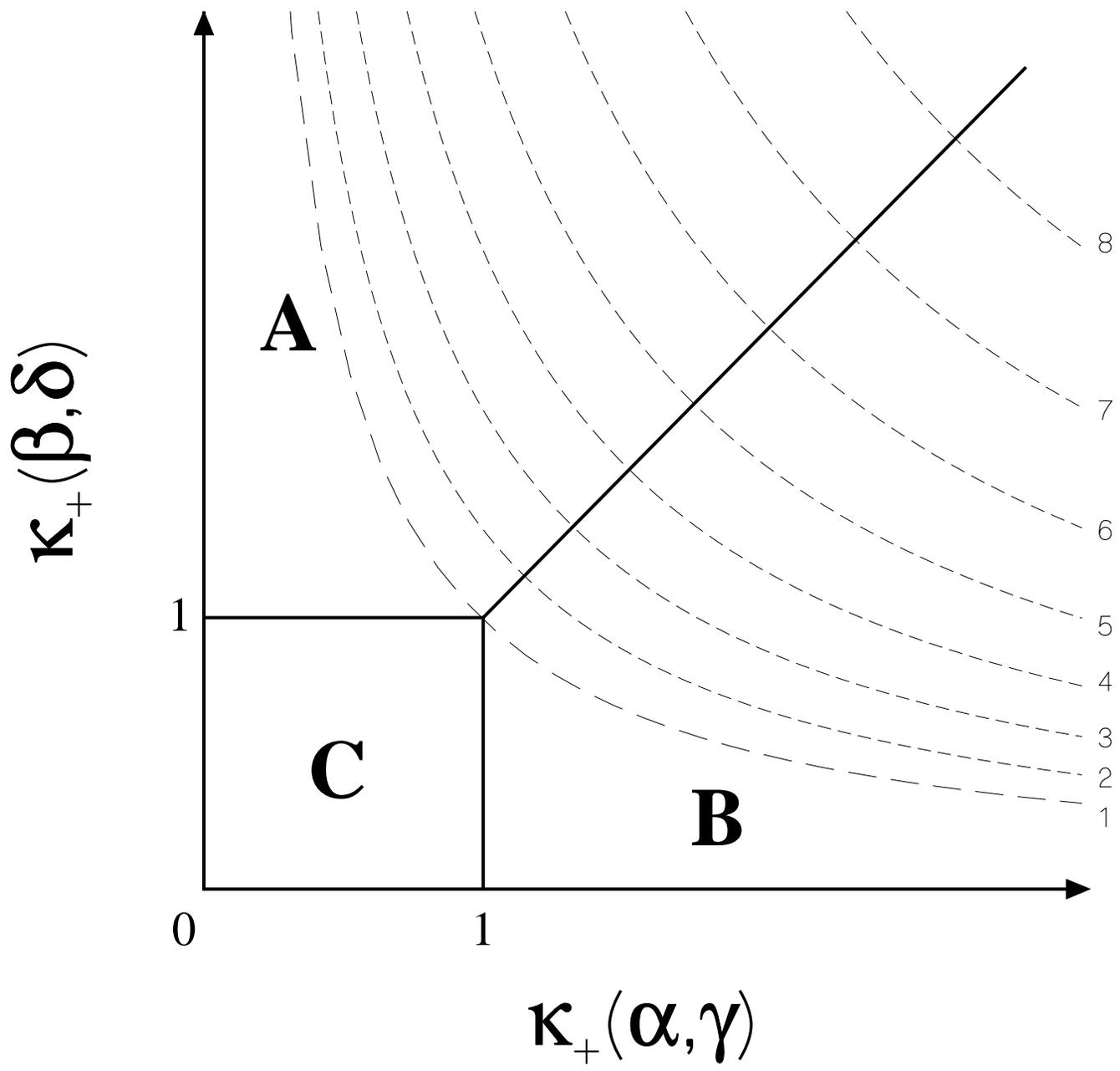
\end{document}